\documentclass[12pt]{article} 
\date{}
\usepackage{amssymb} 
\usepackage{amsmath} 
\def\pd#1#2{\frac{\partial#1}{\partial#2}}      

\def\oint{{\relax \int\kern -1. em O}}

\begin{document}

\title{\bf Riccati equation, Factorization Method and Shape Invariance}

\author{Jos\'e F. Cari\~nena and Arturo Ramos}
\maketitle

\begin{center}
Departamento de F\'{\i}sica Te\'orica. Facultad de Ciencias. \\
Universidad de Zaragoza, 50009, Zaragoza, Spain.

\end{center}  

\begin{abstract}
The basic concepts of factorizable problems in one--dimensional
Quantum Mechanics, as well as the theory of Shape Invariant potentials
are reviewed. The relation of this last theory with a generalization
of the classical Factorization Method presented by Infeld and Hull is
analyzed in detail. By the use of some properties of the Riccati equation
the solutions of Infeld and Hull are greatly generalized in a simple way. PACS numbers: 11.30.Pb, 03.65.Fd.
\end{abstract}

\def\ba{\begin{eqnarray}}
\def\ea{\end{eqnarray}}
\newcommand{\eq}{\begin{equation}}
\newcommand{\feq}{\begin{equation}}
\def\Eq#1{{\begin{equation} #1 \end{equation}}}
\def\R{\mathbb R}
\def\C{\mathbb C}
\def\Z{\mathbb Z}
\def\a{\alpha}                  
\def\b{\beta}                   
\def\g{\gamma}                  
\def\d{\delta}                  

\def\la#1{\lambda_{#1}}
\def\teet#1#2{\theta [\eta _{#1}] (#2)}
\def\tede#1{\theta [\delta](#1)}
\def\N{{\frak N}}
\def\GR{{\cal G}}
\def\Wei{\wp}
\newcommand{\bea}{\begin{eqnarray}}
\newcommand{\eea}{\end{eqnarray}}

\font\tengoth=eufm10 \font\sevengoth=eufm7 \font\fivegoth=eufm5
\newfam\gothfam
\textfont\gothfam=\tengoth \scriptfont\gothfam=\sevengoth
  \scriptscriptfont\gothfam=\fivegoth
  \def\goth{\fam\gothfam}               
\font\frak=eufm10 scaled\magstep1

\def\bra#1{\langle#1|}
\def\ket#1{|#1\rangle}
\def\goth #1{\hbox{{\frak #1}}}
\def\<#1>{\langle#1\rangle}
\def\cotg{\mathop{\rm cotg}\nolimits}
\def\Map{\mathop{\rm Map}\nolimits}
\def\wt{\widetilde}
\def\const{\hbox{const}}
\def\grad{\mathop{\rm grad}\nolimits}
\def\Div{\mathop{\rm div}\nolimits}
\def\braket#1#2{\langle#1|#2\rangle}
\def\Erf{\mathop{\rm Erf}\nolimits}
\def\matriz#1#2{ \left( \begin{array}{#1}{#2} \end{array} \right)}
\def\Eq#1{{\begin{equation} #1 \end{equation}}}
\def\deter#1#2{\left| \begin{array}{#1} {#2} \end{array} \right| }
\def\pd#1#2{\frac{\partial#1}{\partial#2}}
\def\til{\tilde}

\newtheorem{theorem}{Theorem}
\newtheorem{corollary}{Corollary}
\newtheorem{proposition}{Proposition}
\newtheorem{definition}{Definition}
\newtheorem{lemma}{Lemma}
\def\Eq#1{{\begin{equation} #1 \end{equation}}}
\def\R{{\mathbb R}}
\def\C{\mathbb C}
\def\Z{\mathbb Z}
\def\d{\partial}


%
%
%
\def\summn{\frac{\sum_{i=1}^n m_i}{n}}
\def\summj{\sum_{i=1}^n m_j}
\def\suc{\sum_{i=1}^n c_i}
\def\sumc{c_0+\sum_{i=1}^n m_i c_i}
\def\sumcj{c_0+\sum_{j=1}^n m_j c_j}
\def\Dtil{D_0+\sum_{i=2}^n D_i (m_i-m_1)}

\newenvironment{tab}{\fontsize{8}{17}\selectfont}{}
\def\bt{\begin{tab}}
\def\et{\end{tab}}
\def\cps{\ensuremath{b=(b_1,\,\dots,\,b_n)}}    
\def\cp{\ensuremath{b}}                         

\vskip 1cm

\section{Introduction}

The Factorization Method introduced by Schr\"odinger \cite{Sch1,Sch2,Sch3}
and later developed by Infeld and Hull \cite{InfHul} has been 
shown to be very efficient in the search of exactly solvable 
potentials and his interest has been increasing 
since the introduction by Witten of Supersymmetric 
Quantum Mechanics (SUSY)\cite{Witten81}.
The bridge between the theory of solvable potentials in one dimension
and SUSY was established by  
Gendenshte\"{\i}n \cite{Gen} who introduced the concept of 
a discrete reparametrization invariance, usually 
called \lq\lq shape invariance''. 
When studying all these related subjects one is really 
wondered by the almost complete ubiquity  
of some specific Riccati equations appearing in the theory.  

The Riccati equation, which is the simplest first order 
nonlinear differential equation, has a close relation 
with the group $SL(2,\R)$ in the sense established in the celebrated, 
and unfortunately not so well known as it worths, 
Lie Scheffers theorem \cite{LS}. This theorem 
characterizes those first order differential equation systems 
admitting a nonlinear superposition principle. 
It is also known that the problem of finding the general solution 
for these systems is simplified from the knowledge 
of one or more particular solutions. 
All these aspects have been studied from a group theoretical
perspective \cite{CarMarNas,CarRamGra} with special emphasis
in the Riccati equation \cite{CarRam}, which is nothing but the
simplest prototype of equation having a nonlinear superposition principle
(apart from the inhomogeneous linear equations, whose superposition
principle reduces to a \emph{linear} one).  

We feel that an appropriate use of the mathematical 
properties of the Riccati equation may be very useful 
in order to obtain a deep insight in the theory of factorizable
problems in Quantum Mechanics, as well as in its particular 
class given by Shape Invariant partner Hamiltonians.

To begin with, the mentioned properties can be used
to obtain a simpler but more complete 
presentation, as well as a better understanding of
the classical results given in \cite{InfHul}.
Indeed, we will prove that such results can be generalized by simply
considering the general solution of certain Riccati equation instead of
particular ones. In the end, all of the obtained solutions will give rise
to specific, but rather general classes of Shape Invariant potentials 
in the sense of \cite{Gen}.

Moreover, the techniques to be developed here
can be very useful for attacking other still unsolved problems.
For instance, one could consider the study of Shape Invariant potentials 
depending on several parameters transformed by translations 
as proposed in \cite{CoopGinKha}, which is the main subject 
of \cite{CarRamtres}.

The paper is organized as follows: In Section 2 we review the theory of 
related operators and establish the concepts of partner potentials and
Shape Invariant ones depending on an arbitrary set of parameters. 
In Section 3 we establish explicitly 
the equivalence between a slight generalization 
of the classical Factorization Method \cite{InfHul}
and the theory of Shape Invariance. 
Section 4 is devoted to the study of an interesting 
differential equation system of key importance in the development
of the subject. The first of its equations is a constant coefficients 
Riccati one for which we will find  the solutions in full generality. 
We will use all these results in Section 5, where
we obtain some rather general classes of factorizable problems which
contain as particular cases the classical  results of \cite{InfHul}. In addition,
these will give rise to several important families of Shape Invariant
potentials which depend on one parameter transformed by translation.

\section{Hamiltonians related by first--order differential
operators. The concept of Shape Invariance.}

The problem of finding related operators having very similar 
spectra is now a well 
established  subject (see  e.g.  \cite{CarMarPerRan} and
 references therein).
Two linear differential
operators $\til H$ and $H$ are said to
be $A$--related  if there exists an operator $A$ such
that $AH=\til H A$, where $A$ need not to be invertible.

Furthermore, if we  assume that the operator $A$ relating $H$
and $\til H$ is a first order differential operator, say,
of the form
\begin{equation}
A=\frac{d}{dx}+W(x)\,,
\label{defA}
\end{equation}
then, the relation $AH=\til H A$, with
\begin{equation}
H=-\frac{d^2}{dx^2}+V(x)\,,\ \ \ \til H=-\frac{d^2}{dx^2}+\til V(x)\,,
\label{defHHtil}
\end{equation}
leads to
\ba
W(V-\til V)=-W''-V'\,,\quad V-\til V=-2 W'\,,     \label{1stset}
\ea
while the relation $HA^{\dagger}=A^{\dagger}\til H$
leads to
\ba
W(V-\til V)=W''-\til V'\,,\quad V-\til V=-2 W'\,.\label{2ndset}
\ea
One can easily integrate both pair of equations; from the
first pair we obtain the equation $-2 W W'=-W''-V'$ and therefore
$$
V=W^2-W'+c\,,
$$
with $c$ being an integrating constant. Following 
the same pattern with the second pair we have
$$
\til V=W^2+W'+d\,,
$$
$d$ being also a constant. But taking into account 
$V-\til V=-2 W'$ we have $c=d$. We have then the important
property that two Hamiltonians $H$ and $\til H$ of the form
given by (\ref{defHHtil}) can be related by a first order
differential operator $A$ given by (\ref{defA}) if and only if
there exists a real constant $d$ such that $W$ satisfies the
pair of Riccati equations
\ba
V-d&=&W^2-W'\,,         \label{ricV}  \\
\til V-d&=&W^2+W'\,.    \label{ricVtil}  
\ea
Moreover, this means that both Hamiltonians can be factorized
as
\begin{equation}
H=A^\dagger A+d\,,\ \ \ \til H=A A^\dagger+d\,.
\end{equation} 
Adding and subtracting equations (\ref{ricV}) and (\ref{ricVtil})
we obtain the equivalent pair which relates $V$ and $\til V$
\ba
\til V-d&=&-(V-d)+2W^2\,,       \label{relVVtilcuad} \\
\til V&=&V+2 W'\,.              \label{relVVtilder} 
\ea
The potentials  $\til V$ and $V$ are usually  said to be
\emph{partners}.
An important concept is the so called \emph{Shape Invariance} introduced by 
Gendenshte\"{\i}n \cite{Gen}.  He  supposed that $V$ did 
depend on a certain
set of parameters and considered the equations (\ref{ricV}) and (\ref{ricVtil})
as a definition of $V$ and $\til V$  in terms of a \emph{superpotential} $W$.
After, he asked himself what condition was necessary in order to get 
a partner  $\til V$ of the same form as $V$ but for a different choice 
of the values of the parameters involved in $V$.
This relation between $V$ and $\til V$  is now 
commonly known as Shape Invariance of the potentials \cite{Gen}.  

More explicitly, we will suppose that our potentials are
$V=V(x,a)$ and $\til V=\til V(x,a)$, where $a$ denotes a set of parameters.
Gendenshte\"{\i}n \cite{Gen} showed that if we assume the further relation
between $V(x,a)$ and $\til V(x,a)$ given by
\begin{equation}
\til V(x,a)=V(x,f(a))+R(f(a))\,,
\label{SIGed}
\end{equation}
where $f$ is an (invertible and differentiable) transformation 
over the set of parameters $a$, then the complete spectrum of
the Hamiltonians $H$ and $\til H$ can be found easily.
 Just writing the $a$--dependence
the equations (\ref{ricV}), (\ref{ricVtil}) become  
\ba
V(x,a)-d&=&W^2-W'\,,            \label{ricVSI}  \\
\til V(x,a)-d&=&W^2+W'\,.       \label{ricVtilSI}  
\ea

The simplest way of satisfying these equations 
is assuming that $V(x,a)$ and $\til V(x,a)$ are 
obtained from a superpotential function $W(x,a)$ by means of
\ba
V(x,a)-d&=&W^2(x,a)-W'(x,a)\,,                  \label{ricVSIsp}  \\
\til V(x,a)-d&=&W^2(x,a)+W'(x,a)\,.             \label{ricVtilSIsp}  
\ea
The Shape Invariance property in the sense of
\cite{Gen} requires the further condition (\ref{SIGed}) to be
satisfied.

Let us remark that the parameter  $a$ as well as the
transformation law $f(a)$ are completely arbitrary up to now, 
apart from natural requirements as differentiability and 
invertibility.
It is clear that the election of $a$ 
and $f(a)$ is what defines the different classes of Shape Invariant
potentials. In principle, there is no reason why the intersection of 
these classes should be empty. We will consider a simple but important 
type in Section~\ref{InHuSI}.

\section{Equivalence between Shape Invariant potentials
and the Factorization Method\label{eqfmsi}}

We consider in this section a slight generalization of the Factorization
Method as appeared in the celebrated paper \cite{InfHul}. We will prove 
its equivalence with the theory of Shape Invariant partner potentials in the 
sense of \cite{Gen}.

Then, we will deal with the problem of factorizing the linear second--order
ordinary differential equation
\begin{equation}
\frac{d^2 y}{dx^2}+r(x,a)y+\lambda y=0\,,
\label{SODE_gen}
\end{equation}
where the symbol $a$ denotes a set of $n$ independent real parameters, that is,
$a=(a_1,\,\dots,\,a_n)$.
Let us consider  a transformation on  such parameter space
 $f(a)=(f_1(a),\,\dots,\,f_n(a))$.
We will denote by $f^k$, where $k$ is a positive integer, the 
composition of $f$ with itself $k$ times. For a  negative integer $k$ we
will consider the composition of $f^{-1}$ with itself $k$ times
and $f^0$ will be the identity.
The admissible values of the parameters will be $f^{l}(a)$, where $l$ is
an integer restricted to some subset to be precised later. 
The number $\lambda$ is in principle the eigenvalue to be determined. 

In a similar way as in \cite{InfHul},
we will say that (\ref{SODE_gen}) can be factorized if it 
can be replaced by each  of the two following equations:
\ba
H_{+}^{f^{-1}(a)}\,H_{-}^{f^{-1}(a)}y(\lambda,a)
&=&[\lambda-L({f^{-1}(a)})]y(\lambda,a)\,, \label{fac1_gen}         \\
H_{-}^{a}\,H_{+}^{a}y(\lambda,a)
&=&[\lambda-L(a)]y(\lambda,a)\,,
\label{fac2_gen}
\ea
where
\begin{equation}
H_{+}^{a}=\frac{d}{dx}+k(x,a)\,, \ \ \ H_{-}^{a}=-\frac{d}{dx}+k(x,a)\,.
\label{defHs_gen}
\end{equation}
Here, $k(x,a)$ is a function to be determined 
which depends on the set of parameters $a$, and $L(a)$ is a real
number for each value of the $n$--tuple $a$. 
The fundamental idea of this  
generalization is expressed in the following

\begin{theorem}
Let us suppose that our differential equation (\ref{SODE_gen})
can be factorized in the previously  defined sense.
If $y(\lambda,a)$ is one of its solutions then
\ba
y(\lambda,{f^{-1}(a)})&=&H_{-}^{f^{-1}(a)}y(\lambda,a)
\label{up_gen} \\
y(\lambda,f(a))&=&H_{+}^{a}y(\lambda,a)
\label{down_gen}
\ea
are also solutions corresponding to the same $\lambda$ but to
different values of the parameter $n$--tuple $a$, as it is 
suggested by the notations.
\end{theorem}

\smallskip
Proof.- Multiplying (\ref{fac1_gen}) by $H_{-}^{f^{-1}(a)}$ and
(\ref{fac2_gen}) by $H_{+}^{a}$ we have
$$
H_{-}^{f^{-1}(a)}\,H_{+}^{f^{-1}(a)}\,H_{-}^{f^{-1}(a)}y(\lambda,a)
=[\lambda-L({f^{-1}(a)})]H_{-}^{f^{-1}(a)}y(\lambda,a)\,,
$$
$$
H_{+}^{a}\,H_{-}^{a}\,H_{+}^{a}y(\lambda,a)
=[\lambda-L(a)]H_{+}^{a}y(\lambda,a)\,.
$$
Comparison of these equations with (\ref{fac1_gen}) and (\ref{fac2_gen})
shows that $y(\lambda,{f^{-1}(a)})$ as 
defined by (\ref{up_gen}) is a solution of (\ref{SODE_gen})
with $a$ replaced by ${f^{-1}(a)}$. Similarly $y(\lambda,f(a))$ given
by (\ref{down_gen})  is a solution with $a$ replaced by $f(a)$.

\medskip 
It is  to be remarked  that  (\ref{up_gen}) or
(\ref{down_gen}) may give rise to the zero  function; actually, we will
see  that this is necessary at some stage in order to obtain a sequence of 
square--integrable wave functions. 

   Indeed we are only interested here in square integrable 
solutions $y(\lambda,a)$. As we are dealing 
with one--dimensional problems, these solutions can be taken
as real functions. Under this domain the following Theorem
holds:
\begin{theorem}
The linear operators $H_{+}^{a}$ and $H_{-}^{a}$ are formally  
mutually adjoint. That is, if $\phi\psi$ vanishes at 
the ends of the interval $I$,
\begin{equation}
\int_I \phi (H_{-}^{a}\psi)\,dx=\int_I \psi(H_{+}^{a}\phi)\,dx\ .
\end{equation}
\label{mut_adj}
\end{theorem}
\smallskip
Proof.- It is proved directly:
\ba
&& \int_I \phi (H_{-}^{a}\psi)\,dx
=-\int_I \phi\,\frac {d\psi} {dx}\,dx+\int_I \phi\,k(x,a)\,\psi\,dx
                                                                \nonumber\\
&&\qquad\qquad\quad
=\int_I \psi\,\frac {d\phi} {dx}\,dx+\int_I \phi\,k(x,a)\,\psi\,dx
=\int_I \psi(H_{+}^{a}\phi)\,dx\,,                               \nonumber      
\ea
where we have integrated the first term by parts and used  that
$\psi\phi_{|\partial I}=0$.
\smallskip

 Moreover, it is important to know when  (\ref{up_gen})
and (\ref{down_gen}) produce new square--integrable functions.
\begin{theorem}
Let $y(\lambda,a)$ be a non--vanishing, square--integrable  
solution of (\ref{fac1_gen}) and (\ref{fac2_gen}).
The solution $y(\lambda,{f^{-1}(a)})$ defined by (\ref{up_gen}) is
square--integrable if and only if $\lambda\geq L(f^{-1}(a))$. Similarly,
the solution $y(\lambda,{f(a)})$ defined by (\ref{down_gen}) is
square--integrable if and only if $\lambda\geq L(a)$. 
\label{sq_int}
\end{theorem}
\smallskip
Proof.- It is sufficient to compute
\ba
&& \int_I y(\lambda,{f^{-1}(a)})^2\,dx
=\int_I H_{-}^{f^{-1}(a)}y(\lambda,a) H_{-}^{f^{-1}(a)}y(\lambda,a)\,dx \nonumber\\
&&=\int_I y(\lambda,a) (H_{+}^{f^{-1}(a)}H_{-}^{f^{-1}(a)}y(\lambda,a))\,dx
=(\lambda-L({f^{-1}(a)}))\int_I y(\lambda,a)^2 \,dx\,,          \nonumber
\ea
where it has been used Theorem~\ref{mut_adj} and (\ref{fac1_gen}). 
In a similar way,
\ba
&& \int_I y(\lambda,{f(a)})^2\,dx
=\int_I H_{+}^{a}y(\lambda,a) H_{+}^{a}y(\lambda,a)\,dx
                                                        \nonumber\\
&&=\int_I y(\lambda,a) (H_{-}^{a}H_{+}^{a}y(\lambda,a))\,dx
=(\lambda-L(a))\int_I y(\lambda,a)^2 \,dx\,,    \nonumber
\ea
where use has been made of Theorem~\ref{mut_adj} and (\ref{fac2_gen}).

\smallskip

We will consider now the sequence $L(f^k(a))$ and analyze  only the  
cases where it is either an increasing or a decreasing
sequence.  A more complicated behavior of 
$L(f^k(a))$ with $k$ (e.g. oscillatory) will not be treated here. 

\begin{theorem}
Suppose that $L(f^k(a))$ is a decreasing sequence with no accumulation points. Then
the necessary and sufficient condition for having square--integrable 
solutions of the equations (\ref{fac1_gen}) and (\ref{fac2_gen}) is that
there exists a point of the   parameter space, 
$\cps$, such that 
\begin{equation}
\lambda=L(\cp)\,,\quad H_{-}^{\cp}y(\lambda,f(\cp))=0\,, 
\end{equation}
provided the function $y(L(\cp),f(\cp))$ so obtained is square--integrable.
\label{L_dec}
\end{theorem}
\smallskip
Proof.- Let $y(\lambda,a)$ be a non--vanishing, square--integrable  
solution of (\ref{fac1_gen}) and (\ref{fac2_gen}). In order to avoid
a contradiction it is necessary, by Theorem~\ref{sq_int},
that $\lambda\geq L(f^{-1}(a))$. If the 
equality does not hold, one can iterate the process to obtain
\ba
\int_I y(\lambda,{f^{-2}(a)})^2\,dx
=(\lambda-L({f^{-2}(a)}))
(\lambda-L({f^{-1}(a)}))\int_I y(\lambda,a)^2 \,dx\,.           \nonumber
\ea 
Since $L(f^k(a))$ is decreasing with $k$, we have that the difference 
$\lambda-L({f^{-2}(a)})$ is positive or vanishing and smaller 
than $\lambda-L({f^{-1}(a)})$. If it still does not vanish, 
the process can be
continued until we arrive at a value $k_0$ such that
$\lambda=L({f^{-k_0}(a)})$. It is then necessary 
that $y(\lambda,f^{-k_0}(a))=H_{-}^{f^{-k_0}(a)}y(\lambda,f^{-k_0+1}(a))=0$.
It suffices to set $\cp=f^{-k_0}(a)$ to obtain the result.
\smallskip

\begin{theorem}
If $L(f^k(a))$ is an increasing sequence with no accumulation points, then
the necessary and sufficient condition for having square--integrable 
solutions of the equations (\ref{fac1_gen}) and (\ref{fac2_gen}) is that
there exists a specific point of the parameter space, $\cps$,
 such that 
\begin{equation}
\lambda=L(\cp)\,, \quad H_{+}^{\cp}y(\lambda,\cp)=0\,,
\end{equation}
provided the function $y(L(\cp),\cp)$ so obtained is square--integrable.
\label{L_crec}
\end{theorem}
\smallskip
Proof.- Let $y(\lambda,a)$ be a non--vanishing, square--integrable  
solution of (\ref{fac1_gen}) and (\ref{fac2_gen}). In order to avoid
a contradiction it is necessary by Theorem~\ref{sq_int}
that $\lambda\geq L(a)$. If the 
equality does not hold, one can iterate the process to obtain
\ba
\int_I y(\lambda,{f^{2}(a)})^2\,dx
=(\lambda-L(f(a)))
(\lambda-L(a))\int_I y(\lambda,a)^2 \,dx\,.             \nonumber
\ea 
Since $L(f^k(a))$ is an increasing sequence, 
$\lambda-L(f(a))$ is positive or vanishing and smaller 
than $\lambda-L(a)$. If it still does not vanish, 
the process can be
continued until we arrive at $k_0$ such that
$\lambda=L(f^{k_0-1}(a))$. Then, it is necessary  
$y(\lambda,f^{k_0}(a))=H_{+}^{f^{k_0-1}(a)}y(\lambda,f^{k_0-1}(a))=0$.
It suffices to set $\cp=f^{k_0-1}(a)$. 
\medskip
 
When $L(f^k(a))$ is a decreasing 
(resp. increasing) sequence, the functions $y$ 
defined by $H_{-}^{\cp}y(L(\cp),f(\cp))=0$ (resp. 
$H_{+}^{\cp}y(L(\cp),\cp)=0$), provided they are square--integrable,
will be those from where all the others will be constructed. 

We consider now what
relation  among  $r(x,a)$, $k(x,a)$ and $L(a)$ there exists.
Carrying out explicitly
the calculations involved in (\ref{fac1_gen}) and (\ref{fac2_gen}), 
using (\ref{SODE_gen}), we find the equations
\ba
k^2(x,{f^{-1}(a)})+\frac{dk(x,{f^{-1}(a)})}{dx}=-r(x,a)-L({f^{-1}(a)})\,,&&
\label{relkr1_gen} \\
k^2(x,a)-\frac{dk(x,a)}{dx}=-r(x,a)-L(a)\,.\quad\quad&&
\label{relkr2_gen}
\ea
We can eliminate $r(x,a)$ between these equations, obtaining 
\begin{equation}
k^2(x,{f^{-1}(a)})-k^2(x,a)
+\frac{dk(x,{f^{-1}(a)})}{dx}+\frac{dk(x,a)}{dx}=L(a)-L({f^{-1}(a)})\,.
\label{lhs_gen}
\end{equation}
Moreover, since (\ref{relkr1_gen}) and (\ref{relkr2_gen}) 
hold for each $f^{k}(a)$, $k$ in the range of integers corresponding to
square--integrable solutions, we can rewrite them as
\ba
k^2(x,a)+\frac{dk(x,a)}{dx}&=&-r(x,f(a))-L(a)\,, \label{rickrLp_gen}\\  
k^2(x,a)-\frac{dk(x,a)}{dx}&=&-r(x,a)-L(a)\,,   \label{rickrLm_gen}
\ea
and from them we can obtain the equivalent pair
\ba
&& r(x,a)+r(x,f(a))+2\,k^2(x,a)+2\,L(a)=0\,,
                                                \label{si1_gen} \\
&& \quad r(x,a)-r(x,f(a))-2\,\frac{dk(x,a)}{dx}=0\,.
                                                \label{si2_gen}
\ea
Both of the equations (\ref{relkr1_gen}) and (\ref{relkr2_gen})
are necessary conditions  to be satisfied by $k(x,a)$ and $L(a)$, 
for a given $r(x,a)$. They are also sufficient since any $k(x,a)$ and $L(a)$ 
satisfying these equations lead unambiguously 
to a function $r(x,a)$ and so to
a problem whose factorization is known. It should be noted, however,
that there exists the possibility that
equations (\ref{relkr1_gen}) and (\ref{relkr2_gen}) did not have
in general a unique solution for $k(x,a)$ and $L(a)$ for a given $r(x,a)$.

The equation (\ref{lhs_gen}) is what one uses in practice in order to obtain 
results of the Factorization Method.
We try to solve (\ref{lhs_gen}) instead of (\ref{relkr1_gen}) and 
(\ref{relkr2_gen}) since is easier
to find  problems which are factorizable by construction 
than seeing whether certain problem defined by some $r(x,a)$ is factorizable or not.  

Conversely, a solution $k(x,a)$ of
(\ref{lhs_gen}) gives rise to unique expressions for the differences 
$-r(x,f(a))-L(a)$ and $-r(x,a)-L(a)$ by means of equations (\ref{rickrLp_gen}) 
and (\ref{rickrLm_gen}), but it \emph{does not} determine the quantities $r(x,a)$
and $L(a)$ in a unique way. In fact, the method does
 not determine the function 
$L(a)$ unambiguously but only the difference $L(f(a))-L(a)$. And this 
does not define $L(a)$ in a unique way at all. To begin with, $L(a)$ is always 
defined up to a constant. And more ambiguity could arise in
 some cases, as it happens
 in the case studied in \cite{CarRamtres}. But for the purposes
of the application of this method to Quantum Mechanics the interesting quantity
is $L(f(a))-L(a)$, as we will see below. 
The same way is undetermined $r(x,a)$,
with an ambiguity which cancels out exactly with that of $L(a)$ since the
differences $-r(x,f(a))-L(a)$ and $-r(x,a)-L(a)$ are completely determined 
 from a given solution $k(x,a)$ of (\ref{lhs_gen}). 

Going back to the problem of finding 
Shape Invariant potentials in the sense of \cite{Gen} which
depend on the same set of parameters $a$, we remember that
the equations to be satisfied are (\ref{ricVSIsp}) and  (\ref{ricVtilSIsp})
or the equivalent equations 
\ba
\til V(x,a)-d&=&-(V(x,a)-d)+2\,W^2(x,a)\,,      \label{relVVtilcuadSI_gen} \\
\til V(x,a)&=&V(x,a)+2\,W'(x,a)\,,              \label{relVVtilderSI_gen} 
\ea
as well as the Shape Invariance condition (\ref{SIGed}).

Remember that the potentials $V(x,a)$ and $\til V(x,a)$ define
a pair of Hamiltonians
\begin{equation}
H(a)=-\frac{d^2}{dx^2}+V(x,a)\,,\quad
\til H(a)=-\frac{d^2}{dx^2}+\til V(x,a)\,,
\end{equation}
which can be factorized as 
\begin{equation}
H(a)=A(a)^\dagger A(a)+d\,,\quad\quad\til H(a)=A(a) A(a)^\dagger+d\,,
\label{Ham_as}
\end{equation} 
where $d$ is a real number and 
\begin{equation}
A(a)=\frac{d}{dx}+W(x,a)\,,\quad\quad A^\dagger(a)=-\frac{d}{dx}+W(x,a)\,.
\label{defAa}
\end{equation} 
The Shape Invariance condition reads in terms of these Hamiltonians 
\begin{equation}
\til H(a)=H(f(a))+R(f(a))\,.
\label{SIGed_gen_Ham}
\end{equation}

We establish next the identifications between the symbols used in the
generalized  Factorization Method treated in this section and those
used in the theory of Shape Invariance. We will see that the equations
to be satisfied are exactly the same, and that both problems essentially
coincide when we consider square--integrable solutions.
For that purpose is sufficient to identify  
\ba
\til V(x,a)-d&=&-r(x,f(a))-L(a)\,,              \label{idVtilrL_gen}    \\
V(x,a)-d&=&-r(x,a)-L(a)\,,                      \label{idVrL_gen}       \\ 
W(x,a)&=&k(x,a)\,,                              \label{idWk_gen}        \\
R(f(a))&=&L(f(a))-L(a)\,,                       \label{idRL_gen}
\ea
and as an immediate consequence, 
\begin{equation}
A(a)=H_{+}^a\,,\quad\quad A^\dagger(a)=H_{-}^a\,,
\label{id_Hs_As}
\end{equation} 
for all allowed values of $a$.
In fact, with these identifications it is immediate to see  
that equations 
(\ref{rickrLp_gen}) and (\ref{rickrLm_gen}) 
are equivalent
to (\ref{ricVtilSIsp}) and (\ref{ricVSIsp}), respectively.
Moreover
\ba
\til V(x,a)-V(x,f(a))&=&-r(x,f(a))-L(a)+r(x,f(a))+L(f(a))       \nonumber\\
&=&L(f(a))-L(a)=R(f(a))\,,                                      \nonumber
\ea
which is nothing but equation (\ref{SIGed});  
equations (\ref{si1_gen}), (\ref{si2_gen}) become
\ba
&&-(V(x,a)-d)-L(a)-(\til V(x,a)-d)-L(a)+2\,W^2(x,a)+2\,L(a)     \nonumber\\
&&\ \ \ =-(V(x,a)-d)-(\til V(x,a)-d)+2\,W^2(x,a)=0\,,           \nonumber
\ea
and 
\ba
&&-(V(x,a)-d)-L(a)+(\til V(x,a)-d)+L(a)-2\,W'(x,a)      \nonumber\\
&&\ \ \ =-V(x,a)+\til V(x,a)-2\,W'(x,a)=0\,,            \nonumber
\ea
i.e., equations (\ref{relVVtilcuadSI_gen})
and (\ref{relVVtilderSI_gen}), respectively.

But the identification does not stop here. Let us assume that 
Theorem~\ref{L_dec} is applicable. We shall see what it means in terms of
the Hamiltonians (\ref{Ham_as}). To begin with, we have a 
certain point of the parameter space
$\cps$ such that $\lambda=L(\cp)$ and 
$A^\dagger(\cp)y(L(\cp),f(\cp))=0$, where the function 
$y(L(\cp),f(\cp))$ so defined
is square--integrable. We will omit its first argument for brevity, writing
$y(f(\cp))$. It is given by the expression
\begin{equation}
y(f(\cp))=N \exp\left({\int^x W(\xi,\cp)\,d\xi}\right)\,,
\label{est_f_L_dec}
\end{equation}
where $N$ is a normalization constant. Note that this wave function has no 
nodes.
Since $L(f^k(a))$ is a decreasing sequence, we have that
the function $R(f^k(\cp))=L(f^k(\cp))-L(f^{k-1}(\cp))<0$ for all 
of the acceptable values of $k$.

Then, it is easy to check that $y(f(\cp))$ is the ground state of the
 Hamiltonian
$\til H(\cp)$, with energy  $d$. In fact,
$$
\til H(\cp)y(f(\cp))=(A(\cp) A(\cp)^\dagger+d)y(f(\cp))=d\,y(f(\cp))\,.
$$
 From equation (\ref{SIGed_gen_Ham}) we have 
$H(\cp)=\til H(f^{-1}(\cp))-R(\cp)$. The function $y(\cp)$ is the
ground state of $H(\cp)$ with energy $d-R(\cp)$:
$$
H(\cp)y(\cp)=\til H(f^{-1}(\cp))y(\cp)-R(\cp)y(\cp)=(d-R(\cp))y(\cp)\,.
$$
Now, the first excited state of $\til H(\cp)$ is $A(\cp)y(\cp)$:
$$
\til H(\cp)A(\cp)y(\cp)=A(\cp)H(\cp)y(\cp)=(d-R(\cp))A(\cp)y(\cp)\,,
$$
where it has been used the property $\til H(\cp)A(\cp)=A(\cp)H(\cp)$.
In a similar way it can be proved that $A(f^{-1}(\cp))y(f^{-1}(\cp))$
is the first excited state of $H(\cp)$, with energy 
$d-R(\cp)-R(f^{-1}(\cp))$. One can iterate the procedure in order to
solve completely the eigenvalue problem of the Hamiltonians $H(\cp)$ and
$\til H(\cp)$. The results are summarized in Table~\ref{eig_prob_L_dec}.
Note that $d$ has the meaning of the reference energy chosen for the 
Hamiltonians. It is usually taken as zero.

\begin{table}
\begin{tabular}{|p{1.35in}|p{1.75in}|p{1.75in}|}
\hline
\multicolumn{1}{|c|}{\bt Eigenfunctions and energies\et}
    &\multicolumn{1}{|c|}{\bt$\til H(\cp)$\et}
     &\multicolumn{1}{|c|}{\bt$H(\cp)$\et}                      \\
\hline
                &               &                               \\    
\bt \quad\quad Ground state\et     
 &\bt\quad\quad\quad$y(f(\cp))$\et&\bt\quad\quad\quad$y(\cp)$\et          \\
                &               &                               \\
 &\bt\quad\quad\quad$d$\et      &\bt\quad\quad\quad$d-R(\cp)$\et          \\
                &               &                               \\
\hline
                &               &                               \\       
\bt\quad\quad$k^{\mbox{th}}$ excited state\et 
         &\bt$A(\cp)\cdots A(f^{-k+1}(\cp))y(f^{-k+1}(\cp))$\et
      &\bt$A(f^{-1}(\cp))\cdots A(f^{-k}(\cp))y(f^{-k}(\cp))$\et        \\
                &               &                               \\
        &\bt$d-\sum_{r=0}^{k-1}R(f^{-r}(\cp))$\et               
        &\bt$d-\sum_{r=0}^{k}R(f^{-r}(\cp))$\et                 \\
                &               &                               \\        
\hline
\end{tabular}
\caption{Eigenfunctions and eigenvalues of $\til H(\cp)$ 
and $H(\cp)$ when Theorem~\ref{L_dec} is applicable. The function
$y(f(\cp))$ is defined by the relation 
$A^\dagger(\cp)y(f(\cp))=0$.} 
\label{eig_prob_L_dec}
\end{table}

A similar pattern can be followed when it is applicable the 
Theorem~\ref{L_crec}, that is, when $L(f^k(a))$ is an increasing sequence.
The results are essentially the same as when the sequence is decreasing
but where now the Hamiltonian with a lower ground 
state energy is $H(\cp)$. The basic square--integrable 
eigenfunction $y(\cp)$ is defined now by $A(\cp)y(\cp)=0$, that is,
\begin{equation}
y(\cp)=M \exp\left({-\int^x W(\xi,\cp)\,d\xi}\right)\,,
\label{est_f_L_crec}
\end{equation}
where $M$ is the normalization constant.
Moreover, now $R(f^k(\cp))>0$
for all of the acceptable values of $k$. The results are 
summarized in Table~\ref{eig_prob_L_crec}. Again, $d$ sets the energy 
reference level of the Hamiltonians.

\begin{table}
\begin{tabular}{|p{1.35in}|p{1.75in}|p{1.75in}|}
\hline
\multicolumn{1}{|c|}{\bt Eigenfunctions and energies\et}
    &\multicolumn{1}{|c|}{\bt$ H(\cp)$\et}
     &\multicolumn{1}{|c|}{\bt$\til H(\cp)$\et}                 \\
\hline
                &               &                               \\    
\bt \quad\quad Ground state\et     
 &\bt\quad\quad\quad$y(\cp)$\et&\bt\quad\quad\quad$y(f(\cp))$\et          \\
                &               &                               \\
 &\bt\quad\quad\quad$d$\et      &\bt\quad\quad\quad$d+R(f(\cp))$\et       \\
                &               &                               \\
\hline
                &               &                               \\       
\bt\quad\quad$k^{\mbox{th}}$ excited state\et 
         &\bt$A^\dagger(\cp)\cdots A^\dagger(f^{k-1}(\cp))y(f^{k}(\cp))$\et
        &\bt$A^\dagger(f(\cp))\cdots A^\dagger(f^{k}(\cp))y(f^{k+1}(\cp))$\et\\
                &               &                               \\
        &\bt$d+\sum_{r=1}^{k}R(f^{r}(\cp))$\et          
        &\bt$d+\sum_{r=1}^{k+1}R(f^{r}(\cp))$\et                        \\
                &               &                               \\        
\hline
\end{tabular}
\caption{Eigenfunctions and eigenvalues of $H(\cp)$ 
and $\til H(\cp)$  when is applicable Theorem~\ref{L_crec}. The function
$y(\cp)$ is defined by the relation $A(\cp)y(\cp)=0$.} 
\label{eig_prob_L_crec}
\end{table}

In both cases the spectra of both Hamiltonians are exactly the same (with
corresponding eigenfunctions shifted in one step) except for the ground state
of one of them, which has the lowest possible energy. Only one of the 
eigenfunctions, either (\ref{est_f_L_dec}) or (\ref{est_f_L_crec}) may be 
square--integrable. It might happen, however, that neither of these functions 
were so. In such a situation none of the schemes we have developed would 
be of use. The conditions on the function $W(x,b)$ such that one of the 
possible ground states exist are explained e.g. in \cite{GenKriv}. 
Essentially it depends on the asymptotic behavior of 
$\int^x W(\xi,\cp)\,d\xi$ as $x\to\pm\infty$.


In view of all of these identifications the following result is stated

\begin{theorem}
The problem of finding the square integrable solutions of the 
factorization of (\ref{SODE_gen}), given by equations (\ref{fac1_gen})
and (\ref{fac2_gen}), is the same as to solve the discrete eigenvalue problem 
of the Shape Invariant Hamiltonians (\ref{Ham_as})
in the sense of \cite{Gen} which depend on the same set of 
parameters.
\label{equiv_SI_gen_FM}
\end{theorem}

\smallskip

We encourage the reader to compare the results obtained in this section
with the ones in \cite[pp. 24--27]{InfHul}, which have inspired this 
generalization.

Let us consider now the simplest but particularly important case of having
only one parameter whose transformation law is a translation, that is, 
\begin{equation}
f(a)=a-\epsilon\,,\quad\quad\mbox{or}\quad\quad f(a)=a+\epsilon\,,
\label{a_tras}
\end{equation}
where $\epsilon\neq 0$. 
In both cases 
we can normalize the parameter in units of $\epsilon$,
introducing the new parameter
\begin{equation}
m=\frac{a}{\epsilon}\,,\quad\quad\mbox{or}\quad\quad m=-\frac{a}{\epsilon}\,,
\label{m_norm}
\end{equation} 
respectively. In each of these two possibilities
the transformation law reads, 
with a slight abuse of the notation $f$,
\begin{equation}
f(m)=m-1\,.
\label{m_tras}
\end{equation}

Then, the equations to be solved for finding 
Shape Invariant potentials, in the sense of \cite{Gen}, 
depending on one parameter transformed by a translation are 
\ba
V(x,m)-d&=&W^2(x,m)-W'(x,m)\,,                  \label{ricVSI1p}        \\
\til V(x,m)-d&=&W^2(x,m)+W'(x,m)\,,             \label{ricVtilSI1p}  
\ea
or the equivalent equations 
\ba
\til V(x,m)-d&=&-(V(x,m)-d)+2\,W^2(x,m)\,,      \label{relVVtilcuadSI1p} \\
\til V(x,m)&=&V(x,m)+2\,W'(x,m)\,,              \label{relVVtilderSI1p} 
\ea
as well as the Shape Invariance condition
\begin{equation}
\til V(x,m)=V(x,m-1)+R(m-1)\,.
\label{SIGed1p}
\end{equation}

As a particular case of Theorem~\ref{equiv_SI_gen_FM} we have the 
following

\begin{corollary}
The problem of finding all factorizable problems 
following the Factorization Method stated in \cite{InfHul} 
is equivalent to find Shape Invariant potentials
in the sense of \cite{Gen} which depend on one parameter 
transformed by translation. 
\label{equiv_SI1p_FM}
\end{corollary}

\smallskip

The relations among the relevant quantities in both approaches
are given next for completeness,
\ba
\til V(x,m)-d&=&-r(x,m-1)-L(m)\,,               \label{idVtilrL}        \\
V(x,m)-d&=&-r(x,m)-L(m)\,,                      \label{idVrL}           \\ 
W(x,m)&=&k(x,m)\,.                              \label{idWk}            \\
R(m-1)&=&L(m-1)-L(m)\,.                         \label{idRL}
\ea

We would like to remark that the e\-qui\-va\-len\-ce bet\-ween the
Fac\-to\-ri\-za\-tion Method and Shape Invariance has been
first pointed out, to our knowledge, by several authors almost 
ten years ago (see e.g. \cite{Sta,Sta2} and \cite{MonSal}). 
It seems to us 
that most of the authors in these subjects have the feeling 
(or even a more precise knowledge) that such identification exists. 
But we have not seen so far a complete and clear identification in the 
general case where arbitrary set of parameters $a$ and transformation
laws $f(a)$ are involved.
Our aim is just to take a step ahead in the task of clarifying how 
these methods are interrelated since they can be used in more general
situations. An important example of this is obtained when an arbitrary
but finite number of parameters subject to translation 
is involved \cite{CarRamtres}.

\section{General solution of equations $y^2+y'=a$, 
$zy+z'=b$\label{solgyz}}

In this section we will study the general solution of an 
ordinary differential 
equation system which  will appear as the key point in the 
solution of the problems posed in \cite{InfHul}, which we will revisit in
the next section. 

Let us  consider the differential equation system in
 the  variables $y$ and $z$
\ba
y^2+y'&=&a\,,                                   \label{ric_y}\\
y z+z'&=&b\,,                                   \label{lin_z}
\ea 
where $a$ and $b$ are real constants and the prime denotes derivative
respect to $x$. The equation (\ref{ric_y}) is a Riccati equation with 
constant coefficients, meanwhile (\ref{lin_z}) is an  inhomogeneous
linear first order differential equation for $z$, provided the function 
$y$ is known. Recall that the general solution of the inhomogeneous 
linear first order differential equation for $v(x)$
\begin{equation}
\frac{dv}{dx}=a(x)v(x)+b(x)\,,
\end{equation}
can be obtained by means of the formula
\begin{equation}
v(x)=\frac{\int^x b(\xi)\exp\big\{-\int^\xi a(\eta)\,d\eta\big\}\,d\xi+E}
{\exp\big\{-\int^x a(\xi)\,d\xi \big\}}\,, \label{sol_gen_lin}
\end{equation}
where $E$ is an integration constant.
Then, the  general solution of (\ref{lin_z}) is 
easily obtained  once we know the solutions of (\ref{ric_y}), i.e. 
\begin{equation}
z(x)=\frac{b\,\int^x\exp\big\{\int^\xi y(\eta)\,d\eta\big\}\,d\xi+D}
{\exp\big\{\int^x y(\xi)\,d\xi \big\}}\,,       \label{sol_z}
\end{equation}
where we name the integration constant as $D$.
So, let us first pay  attention to the task of solving (\ref{ric_y}) in its
full generality. 

The general Riccati equation  
\begin{equation}
\frac{dy}{dx}=a_2(x) y^2+a_1(x) y+a_0(x)\,,  \label{ric_y_gen}
\end{equation}
where $a_2(x)$, $a_1(x)$ and $a_0(x)$ are differentiable functions of 
the independent variable $x$, has very interesting properties. 
We will recall here  some of them which will be of use in our problem.  
It is a non--linear first order differential
equation, and in the most general case there is no way of writing the general
solution by using some quadratures. However, one can integrate it completely
if some extra information is known. 
For example, if one particular solution $y_1(x)$ of (\ref{ric_y_gen}) is 
known, the problem can be reduced to   an inhomogeneous first order linear
equation and the general solution can be found by two quadratures. In fact,
 the change of variable (see e.g. \cite{Dav,Mur})
\begin{equation}
u=\frac{1}{y_1-y}\,,\quad\mbox{with inverse} 
\quad y=y_1-\frac{1}{u}\,,                      \label{ch_1sol_usual}
\end{equation}
transforms (\ref{ric_y_gen}) into the inhomogeneous first order linear
equation
\begin{equation}
\frac{du}{dx}=-(2\,a_2\,y_1+a_1)u+a_2\,,
\end{equation}
which can be integrated by two quadratures, for example using 
(\ref{sol_gen_lin}). An alternative
change of variable was also found in \cite{CarRam}:
\begin{equation}
u=\frac{y\,y_1}{y_1-y}\,,\quad\mbox{with inverse} 
\quad y=\frac{u\,y_1}{u+y_1}\,.   \label{ch_1sol_nues}
\end{equation}
This change transforms (\ref{ric_y_gen}) into the 
inhomogeneous first order linear equation
\begin{equation}
\frac{du}{dx}=\bigg(\frac{2\,a_0}{y_1}+a_1\bigg)u+a_0\,,
    \label{lin_ch_1sol_nues}
\end{equation}
which is integrable by two quadratures, as well. 
 We also remark that the general 
Riccati equation (\ref{ric_y_gen}) admits the identically vanishing 
function as a solution if and only if $a_0(x)=0$ for all $x$. 

Even more interesting is the following property:
once three particular solutions of (\ref{ric_y_gen}), 
$y_1(x),\,y_2(x),\,y_3(x)$, are known, the general solution $y$ can be
written, without making use of any quadrature, by means of the formula
\begin{equation}
\frac{(y-y_1)(y_2-y_3)}{(y-y_2)(y_1-y_3)}=k\ ,
\label{superp_formula}
\end{equation}
where $k$ is a constant determining each solution. 
Solving for $y$ we get
\begin{equation}
y=\frac {y_2(y_3-y_1)\,k+y_1(y_2-y_3)}{(y_3-y_1)\,k+y_2-y_3}\ .
                                                \label{solv_y}
\end{equation}
 As an example, it is easy to check that
$y|_{k=0}=y_1$, $y|_{k=1}=y_3$ and that the solution $y_2$ is obtained as
 the limit of $k$ going to  $\infty$.

The theorem for uniqueness of solutions of differential
equations shows that the difference between two solutions of the Riccati
equation (\ref{ric_y_gen}) has a constant sign and therefore the difference 
between two different solutions never vanishes, and
the quotients in the previous equations are always well defined.

The equation (\ref{solv_y}) furnishes  a non--linear superposition
 principle for the  Riccati equation: there exists a superposition
 function $\Phi(u_1,u_2,u_3,k)$ such that
for any  three particular fundamental solutions, 
the  function $\Phi(y_1,y_2,y_3,k)$  gives  the general 
solution.

The first order differential equation systems having 
this important property are characterized by 
the so called Lie--Scheffers theorem \cite{LS}, the simplest one
being the Riccati equation (apart from the inhomogeneous first order 
linear equation, whose superposition principle 
reduces to a \emph{linear} one).
These problems have had a revival after several interesting papers by 
Winternitz and coworkers (see e.g. \cite{PW1} and references
 therein), 
and have been studied in \cite{CarMarNas} from 
a group theoretical perspective. In \cite{CarRam} the integrability
conditions of the Riccati equation, as well as its non--linear
superposition principle are studied in a unified way by making use of 
an action on the set of Riccati equations. A 
generalization 
to other groups and  systems admitting  such a 
non--linear superposition principle is given in \cite{CarRamGra}.

We are interested here in the simpler case of the
Riccati equation with
constant coefficients (\ref{ric_y}). The general equation of this type is
\begin{equation}
\frac{dy}{dx}=a_2 y^2+a_1 y+a_0\,,  \label{ric_y_const}
\end{equation}
where $a_2$, $a_1$ and $a_0$ are now real constants, $a_2\neq 0$.
For a review of some of its properties from a
geometrical viewpoint see \cite{CarRam}.  
This equation, unlike the general Riccati equation
 (\ref{ric_y_gen}),
is always integrable by quadratures, and the form of the 
solutions depends strongly on the sign of the discriminant 
$\Delta=a_1^2-4a_0 a_2$. This can be seen by separating the differential
equation (\ref{ric_y_const}) in the form
$$
\frac{dy}{a_2 y^2+a_1 y+a_0}=\frac{dy}
{a_2\bigg(\big(y+\frac{a_1}{2\,a_2}\big)^2
-\frac{\Delta}{4\,a_2^2}\bigg)}=dx\,.
$$
Integrating (\ref{ric_y_const}) in this way we obtain non--constant 
solutions. 

Looking for constant solutions of (\ref{ric_y_const}) amounts to solve
an algebraic second order equation. So, if $\Delta>0$ there will be
 two different real constant solutions. If $\Delta=0$ there is
 only one constant 
real solution and if $\Delta<0$ we have no constant
 real solutions at all.

We shall illustrate these properties while  finding the general
solution of (\ref{ric_y}). For this equation 
the discriminant $\Delta$ is just $4a$. 
Then, the form of the solutions depend strongly on the sign of $a$. 
If $a>0$ we can write $a=c^2$, where $c>0$ is a real number. The 
non--constant particular solution
\begin{equation}
y_1(x)=c\tanh(c(x-A))\,,                \label{sp_y1_a>0}
\end{equation}
where $A$ is an arbitrary integration constant, 
is readily found by direct integration. In addition, there exists two
different constant real solutions,
\begin{equation}
y_2(x)=c\,,\quad\quad y_3(x)=-c\,.      \label{scp_y1_a>0}
\end{equation}  
Then, we can find out the general solution from these particular
solutions using
 the non--linear superposition formula (\ref{solv_y}), 
yielding
\begin{equation}
y(x)=c\,\frac{B\,\sinh(c(x-A))-\cosh(c(x-A))}
{B\,\cosh(c(x-A))-\sinh(c(x-A))}\,,
                                                        \label{gen_y_a>0}
\end{equation}
where $B=(2-k)/k$, $k$ being the arbitrary constant 
in (\ref{solv_y}). Substituting in (\ref{sol_z})
we obtain the general solution for $z(x)$,
\ba
z(x)=\frac{\frac{b}{c}\{B\,\sinh(c(x-A))-\cosh(c(x-A))\}+D}
{B\,\cosh(c(x-A))-\sinh(c(x-A))}\,,
\ea 
where $D$ is a new integration constant. 

Let us study now the case with $a=0$ in (\ref{ric_y}).  
By direct integration we find the particular solution
\begin{equation}
y_1(x)=\frac 1{x-A}\,,                          \label{spar_a=0}
\end{equation}
where $A$ is an integration constant. It is clear that now
(\ref{ric_y}) admits the identically vanishing solution, and the 
general solution have to reflect this fact.
In order to find it
is particularly simple the application of the change 
of variable (\ref{ch_1sol_nues}) with $y_1$ given by (\ref{spar_a=0}). 
Indeed, such a change transforms (\ref{ric_y}) with $a=0$  into 
$du/dx=0$, which has the general solution $u(x)=B$, $B$ constant.
Then, the general solution for (\ref{ric_y}) with $a=0$ is
\begin{equation}
y(x)=\frac{B}{1+B(x-A)}\,,
                                                        \label{gen_y_a=0}
\end{equation}
with $A$ and $B$ being arbitrary integration constants. 
If $B=0$ we recover the identically vanishing solution as expected. 
Had we  followed the usual change of variable (\ref{ch_1sol_usual}) 
we would have obtained exactly the same result, but the calculations 
would have been a bit longer.
Substituting in (\ref{sol_z})
we obtain the general solution for $z(x)$ in this case,
\ba
z(x)=\frac{b(\frac B 2 (x-A)^2+x-A)+D}{1+B(x-A)}\,,
\ea 
where $D$ is a new integration constant. 

The last case to be studied is $a<0$. We write then $a=-c^2$,
where $c>0$ is a real number. It is easy to find the non--constant 
particular solution 
\begin{equation}
y_1(x)=-c\tan(c(x-A))\,,                \label{spar_a<0}
\end{equation}
where $A$ is an arbitrary integration constant, by direct integration.
In order to find out the general solution, we make the change of variable
(\ref{ch_1sol_usual}) or alternatively (\ref{ch_1sol_nues}), with $y_1(x)$
given by (\ref{spar_a<0}). In both cases the calculations are essentially 
the same and give the general solution of (\ref{ric_y}) for $a>0$ 
\begin{equation}
y(x)=-c\,\frac{B\,\sin(c(x-A))+\cos(c(x-A))}
{B\,\cos(c(x-A))-\sin(c(x-A))}\,,
                                                        \label{gen_y_a<0}
\end{equation}
where $B=cF$, $F$ an arbitrary constant. 
Substituting in (\ref{sol_z})
we obtain the corresponding general solution for $z(x)$,
\ba
z(x)=\frac{\frac{b}{c}\{B\,\sin(c(x-A))+\cos(c(x-A))\}+D}
{B\,\cos(c(x-A))-\sin(c(x-A))}\,,
\ea 
where $D$ is a new integration constant. 

Needless to say, in all of the three cases the solutions can be written 
in many ways, mostly in the cases where 
exponential, hyperbolic or trigonometric functions are
involved. The choice  of the form in which the arbitrary constants 
appear might also make the solutions to look a bit different, 
but these aspects are irrelevant from the mathematical point of view. 
We have tried to give the simplest form for the solutions and in 
such a way the symmetry between the solutions for the case $a>0$ and
$a<0$ were clearly recognized. Indeed, the general solution of (\ref{ric_y}) 
for $a>0$ can be transformed into that of the case $a<0$ by means of
the formal changes $c\rightarrow ic$, $B\rightarrow iB$ and the identities
$\sinh(ix)=i\sin(x)$, $\cosh(ix)=\cos(x)$. The change for $B$ is motivated
by its definition in the the general solution of (\ref{ric_y}) for $a<0$. 
The results are summarized in Table~\ref{sols_gens}.

We must pay attention to  the following point. If we consider, 
 for instance, the 
general solution of (\ref{ric_y}) for $a>0$, i.e. equation
 (\ref{gen_y_a>0}),
one could be tempted to write it in the form of a logarithmic derivative,
$$
y(x)=\frac{d}{dx}\log|B\,\cosh(c(x-A))-\sinh(c(x-A))|\,,
$$
which is  equivalent except for $B\rightarrow\infty$. 
In fact, if we want to calculate
$$
\lim_{B\to\infty}\frac{d}{dx}\log|B\,\cosh(c(x-A))-\sinh(c(x-A))|
$$
we \emph{cannot} interchange the limit with the derivative,
 otherwise we would get a wrong result. 
The reason, obviously, is that $B\,\cosh(c(x-A))-\sinh(c(x-A))$ is not regular
as $B\to\infty$. But this limit for $B$ is particularly important since 
when taking it in (\ref{gen_y_a>0}), 
we recover the particular solution (\ref{sp_y1_a>0}).
A similar thing happens in the general solutions (\ref{gen_y_a=0}) and
(\ref{gen_y_a<0}), where after 
taking the limit $B\to\infty$ we recover, 
respectively, the particular solutions (\ref{spar_a=0}) and (\ref{spar_a<0})
from which we have started.
Both of (\ref{gen_y_a=0}) and (\ref{gen_y_a<0})
can be written in the form of a logarithmic derivative, but then the limit
$B\to\infty$ could not be calculated properly. 

The conclusion is the following.
If one or more particular solutions of a Riccati equation are known,
the general solution can be found, for example, by one of the methods
described above. This general solution depends on one parameter
characterizing the particular solutions, and in particular one should
be able to recover the known solutions for some specific values.
One of these values is usually infinite. If one writes the 
general solution as a logarithmic derivative, the limit when the
parameter tends to infinite is to be treated with care.

\begin{table}
\begin{tabular}{|p{1in}|p{1.7in}|p{2.2in}|}
\hline
\multicolumn{1}{|c|}{\bt Sign of $a$\et}
        &\multicolumn{1}{|c|}{\bt$y(x)$\et}
                &\multicolumn{1}{|c|}{\bt $z(x)$\et}                   \\
\hline
                &               &                               \\    
\bt\quad$a=c^2>0$\et 
&\bt\quad$c\,\frac{B\,\sinh(c(x-A))-\cosh(c(x-A))}
{B\,\cosh(c(x-A))-\sinh(c(x-A))}$\et
  &\bt\quad\quad$\frac{\frac{b}{c}\{B\,\sinh(c(x-A))-\cosh(c(x-A))\}+D}
  {B\,\cosh(c(x-A))-\sinh(c(x-A))}$\et                          \\
                &               &                               \\
\hline
                &               &                               \\       
\bt\quad$a=0$\et 
                &\bt\quad\quad$\frac{B}{1+B(x-A)}$\et
    &\quad\quad\bt$\frac{b(\frac B 2 (x-A)^2+x-A)+D}{1+B(x-A)}$\et        \\
                &               &                               \\         
\hline
                &               &                               \\        
\bt\quad$a=-c^2<0$\et
  &\bt\quad$-c\,\frac{B\,\sin(c(x-A))+\cos(c(x-A))}
  {B\,\cos(c(x-A))-\sin(c(x-A))}$\et
 &\bt\quad\quad$\frac{\frac{b}{c}\{B\,\sin(c(x-A))+\cos(c(x-A))\}+D}
 {B\,\cos(c(x-A))-\sin(c(x-A))}$\et                             \\
                &               &                               \\
\hline
\end{tabular}
\caption{General solutions of the equations  
(\ref{ric_y}) and (\ref{lin_z}). $A$,
$B$ and  $D$ are integration constants. The constant
$B$ selects the particular solution of (\ref{ric_y}) in each case.}
\label{sols_gens}
\end{table}

\section{The Infeld-Hull Factorization Method revisited: 
Shape Invariant potentials depending
on one parameter transformed by translation\label{InHuSI}}

We will start this section reviewing the steps of the famous 
paper \cite{InfHul}, where the
Factorization Method was developed in a quite systematic way.
It is worth
mentioning, however, that this method take its roots on previous papers by
Schr\"odinger \cite{Sch1,Sch2,Sch3} and others (see references in 
\cite[p. 23]{InfHul}).
We will apply the mathematical theory developed in the
preceding sections for solving the problem in a simple way and with full 
generality, obtaining in the end Shape-Invariant potentials in the sense
of \cite{Gen} depending on one parameter transformed by translation.

The key point in the process of finding factorizable 
problems of type (\ref{SODE_gen})
is to find solutions $k(x,a)$ for the equation (\ref{lhs_gen}),
as we have said in Section~\ref{eqfmsi}. In our current problem 
it takes the form
\begin{equation}
k^2(x,m+1)-k^2(x,m)
+\frac{dk(x,m+1)}{dx}+\frac{dk(x,m)}{dx}=L(m)-L(m+1)\,,
\label{lhs}
\end{equation}
which is a differential--difference equation. The idea of
solving it in its full generality seems to be very difficult,
at least at first sight. Instead of doing that, it seems to
be more sensible to try particular forms of the dependence of
$k(x,m)$ on $x$ and $m$. Then, we should  
find out whether the equation is satisfied in each particular case.

First, note (see \cite{InfHul}) that there exists 
a trivial solution of (\ref{lhs}), namely
$$
k(x,m)=f(m)\,,\ \ \ \ L(m)=-f^2(m)\,,
$$
where $f(m)$ is any function of $m$. This gives rise to the problem
$$
\frac{d^2y}{dx^2}+\lambda y=0\,,
$$
which has been discussed completely by Schr\"odinger \cite{Sch2}.

We next try a solution with an affine dependence on $m$ \cite{InfHul}
\begin{equation}
k(x,m)=k_0(x)+m\,k_1(x)\,,\label{mlin}
\end{equation}
where $k_0$ and $k_1$ are functions of $x$ only. Substituting
into (\ref{lhs}) we obtain the equation
\ba
L(m)-L(m+1)&=&[(m+1)^2(k_1^2+k_1')+2 (m+1) (k_0 k_1+k_0')]      \nonumber\\
                & &-[m^2(k_1^2+k_1')+2 m (k_0 k_1+k_0')]\,.     \label{eqLs}
\ea
Now we would like to reinterpret the reasoning 
followed in \cite[p.27]{InfHul}.
Equation (\ref{eqLs}) reads in its more simplified way
\begin{equation}
L(m)-L(m+1)=2 m (k_1^2+k_1')+k_1^2+k_1'+2 (k_0 k_1+k_0')\,.
\label{eqLssimp}
\end{equation}
Since $L(m)$ is a function of $m$ alone, the coefficients
of the powers of $m$ on the right hand side must be constant. 
Eventually one finds the same coefficients to be constant as in the equation 
appearing after (3.1.4) of \cite{InfHul}. 
Then, the equations to be satisfied are
\ba
k_1^2+k_1'&=&a\,,               \label{eqk1}    \\
k_1 k_0+k_0'&=&b\,,             \label{eqk0}    
\ea  
where $a$ and $b$ are in principle real arbitrary constants.
When these equations are satisfied (\ref{eqLssimp}) becomes 
$$
L(m)-L(m+1)=2(m a+b)+a\,.
$$
We look for the most general polynomial solution of this equation.
It should be of degree two in $m$ if $a\neq 0$ 
(degree one if $a=0$); otherwise
we would find that the coefficients of powers greater 
or equal to three (resp. two) have to vanish. 
Then we put $L(m)=r m^2+s m+t$, where $r,\,s,\,t$ are
constants to be determined. 
Substituting in the previous equation we find the relations
$$
r=-a,\ \ \ \ s=-2\,b,
$$ 
and as a result we have the most general polynomial solution for $L(m)$
\begin{equation}
L(m)=-a m^2-2 b m+t\,,
\end{equation}
where $t$ is an arbitrary real constant. This expression is valid even 
in the case $a=0$, being then $L(m)=-2 b m+t$. 

In \cite[eqs. (3.1.5)]{InfHul} equations (\ref{eqk1}), (\ref{eqk0}) 
are written in the slightly more restricted way (we use Greek characters
to avoid confusion)
\ba
k_1^2+k_1'&=&-\alpha^2\,,            \label{eqk1_InHu}\\
k_1 k_0+k_0'&=&\beta\,,             \label{eqk0_InHu}
\ea  
where $\beta=-\gamma \alpha^2$ if $\alpha\neq 0$. 
This means to consider \emph{only} negative or zero
values of $a$ in (\ref{eqk1}). Indeed, the solutions of
(\ref{eqk1}) for $a>0$ are absent in \cite[eqs. (3.1.7)]{InfHul}, 
which are supposed to be the most general solutions of the system
(\ref{eqk1_InHu}) and (\ref{eqk0_InHu}). However, the solutions
appearing when one considers the solutions of (\ref{eqk1}) for $a>0$
have their own physical importance.
Indeed, Infeld and Hull treat
particular cases of their general factorization types $(A)$, $(B)$ and 
$(E)$ after having made the 
formal change $\alpha\rightarrow -i\alpha$ \cite[pp. 27, 30, 36, 46]{InfHul}. 

But the really important point is that in \cite{InfHul}, even dealing with
their slightly restricted differential equation system (\ref{eqk1_InHu}) 
and (\ref{eqk0_InHu}), they \emph{do not} give the \emph{general} solutions
but simply \emph{particular} ones, since they only consider particular
solutions of the Riccati equation with constant coefficients (\ref{eqk1_InHu}).
They only consider two such solutions  
when $\alpha\neq 0$ and another two when $\alpha=0$.

We would like to point out three main aspects now. First, we will treat the
differential equation system (\ref{eqk1}) and (\ref{eqk0}) for all real values
of $a$ and $b$. We will find the general solutions of the system
by first considering the general solution of the Riccati equation (\ref{eqk1}). 
Second, we will prove that all the solutions
included in the classic paper \cite{InfHul} are particular cases of that
general solutions. Moreover, there is no need of making formal complex changes
of parameters for obtaining some of the relevant physical solutions, 
since they already appear in the general ones. 
Thirdly, we will see that
rather than having four general basic types of factorizable 
problems $(A)$, $(B)$, $(C)$ and $(D)$, where $(B)$, $(C)$ and $(D)$ 
could be considered as limiting forms of $(A)$ \cite[p. 28]{InfHul}, 
there exist indeed three 
general basic types of factorizable problems which include the previously 
mentioned as particular cases, and they are classified by the simple 
distinction of what sign takes $a$ in (\ref{eqk1}). 
The distinction by the sign of $a$ have indeed a deeper geometrical meaning,
but we will not go further in this aspect here. See \cite[Sec. 4]{CarRam}
for more details. 

Moreover, the mentioned lack of generality seems to have been propagated to 
later works trying to generalize
the Factorization Method as exposed in \cite{InfHul}.
See for example some works by  Humi \cite{Hum68,Hum70,Hum86,Hum87}.
There, more general results could be obtained, in principle, by considering
negative values of certain constant appearing in his reasoning and the 
general solution of the Riccati equation which appears rather than particular 
ones. For the last two of these references, it would be necessary to 
consider the general solution of \emph{matrix} Riccati equations, 
which may in turn be formulated by means of certain non--linear 
superposition principle. At  this point, it could be of practical 
use part of the extensive work in the field done by Winternitz and 
coworkers (see e.g. \cite{PW1,PW2,ShnWin} and references therein).  

So, let us find the general solutions of (\ref{eqk1}) and
(\ref{eqk0}). They are just the same as that of the differential
equation system (\ref{ric_y}) and (\ref{lin_z}),
simply identifying $y(x)$ as $k_1(x)$ and $z(x)$ as $k_0(x)$, with the
same notation for the constants. The results are shown 
in Table~\ref{res_p_InHu}.

Next we show how these solutions reduce to the ones contained in 
\cite{InfHul}. 
For the case $a<0$, taking $B\to 0$ we recover
the factorization type $(A)$ of Infeld and Hull \cite[eq. $(3.1.7a)$]{InfHul}.
And taking $B\to i$, with a slight generalization of the values $B$ can take,
we obtain their type $(B)$ (see eq. $(3.1.7b)$). For practical cases of
physical interest, they use these factorization types after
making the formal change $\alpha\to -i\alpha$
\cite[pp. 27, 30, 36, 46]{InfHul}. The same results would be obtained 
if one considers the limiting cases $B\to 0$ or $B\to 1$, respectively,
when $a>0$, so there is no need of making such 
formal changes. For the case $a=0$, taking $B\to \infty$ or 
$B\to 0$ we recover their factorization
types $(C)$ and $(D)$ (see their equations $(3.1.7c)$ and $(3.1.7d)$), 
respectively. 
Remember that our 
convention for the constants 
appearing in equations (\ref{eqk1}) and (\ref{eqk0}) differs slightly 
from that of equations $(3.1.5)$ of \cite{InfHul}, reproduced here as 
(\ref{eqk1_InHu}) and (\ref{eqk0_InHu}) with Greek characters for the 
constants.

We show as well some limiting cases of $B$ which give us the
particular solutions used in the construction of the general ones. 
Remember that the limits $B\to\infty$ should be taken 
with care. The arbitrary constant $D$ appearing in the table is not defined
exactly in the same way in all its occurrences but it always reflects the fact of
having an arbitrary constant wherever it appears.

\begin{table}
\begin{tabular}{|p{.65in}|p{1.5in}|p{2.05in}|p{.5in}|}
\hline
\multicolumn{1}{|c|}{\bt Sign of $a$\et}
        &\multicolumn{1}{|c|}{\bt $k_1(x)$ and limits\et}
           &\multicolumn{1}{|c|}{\bt $k_0(x)$ and limits\et}
                &\multicolumn{1}{|c|}{\bt Comments\et}          \\
\hline
                &               &               &               \\    
\bt$a=c^2>0$\et 
& \bt $c\,\frac{B\,\sinh(c(x-A))-\cosh(c(x-A))}
{B\,\cosh(c(x-A))-\sinh(c(x-A))}$\et
  & \bt $\frac{\frac{b}{c}\{B\,\sinh(c(x-A))-\cosh(c(x-A))\}+D}
  {B\,\cosh(c(x-A))-\sinh(c(x-A))}$\et          &               \\
                &               &               &               \\
 &\bt$\xrightarrow{B \to \infty} c\tanh(c(x-A))$\et             
  &\bt$\xrightarrow{B \to \infty} 
    \frac{b}{c}\tanh(c(x-A))+\frac{D}{\cosh(c(x-A))}$\et
    &\bt See (\ref{sp_y1_a>0})\et                               \\
                &               &               &               \\       
 &\bt$\xrightarrow{B \to 0} c\coth(c(x-A))$\et          
  &\bt$\xrightarrow{B \to 0} 
    \frac{b}{c}\coth(c(x-A))+\frac{D}{\sinh(c(x-A))}$\et
    &\bt See text \et                                                     \\
                &               &               &               \\
 &\bt $\xrightarrow{B \to \mp 1} \pm c$ \et             
  &\bt$\xrightarrow{B \to \mp 1}\pm\frac{b}{c}+D\exp(\mp c(x-A))$\et
    &\bt See (\ref{scp_y1_a>0})\et                              \\       
                &               &               &               \\
\hline
                &               &               &               \\       
\bt$a=0$\et 
                &\bt$\frac{B}{1+B(x-A)}$\et
&\bt$\frac{b(\frac B 2 (x-A)^2+x-A)+D}{1+B(x-A)}$\et &          \\
                &               &               &               \\         
 &\bt$\xrightarrow{B \to \infty} \frac{1}{x-A}$\et              
  &\bt$\xrightarrow{B \to \infty} 
    \frac{b}{2}(x-A)+\frac{D}{x-A}$\et
    &\bt  Type $(C)$\et                                         \\
                &               &               &               \\
 &\bt$\xrightarrow{B \to 0} 0$\et               
  &\bt$\xrightarrow{B \to 0}b(x-A)+D$\et
    &\bt  Type $(D)$\et                                         \\        
                &               &               &               \\
\hline
                &               &               &               \\        
\bt$a=-c^2<0$\et
  &\bt$-c\,\frac{B\,\sin(c(x-A))+\cos(c(x-A))}
  {B\,\cos(c(x-A))-\sin(c(x-A))}$\et
 &\bt$\frac{\frac{b}{c}\{B\,\sin(c(x-A))+\cos(c(x-A))\}+D}
 {B\,\cos(c(x-A))-\sin(c(x-A))}$\et             &               \\
                &               &               &               \\
 &\bt$\xrightarrow{B \to \infty} -c\tan(c(x-A))$\et             
  &\bt$\xrightarrow{B \to \infty} 
    \frac{b}{c}\tan(c(x-A))+\frac{D}{\cos(c(x-A))}$\et
    &\bt See (\ref{spar_a<0})\et                                \\
                &               &               &               \\       
 &\bt$\xrightarrow{B \to 0} c\cot(c(x-A))$\et           
  &\bt$\xrightarrow{B \to 0} 
    -\frac{b}{c}\cot(c(x-A))+\frac{D}{\sin(c(x-A))}$\et
    &\bt Type $(A)$\et                                          \\
                &               &               &               \\
 &\bt$\xrightarrow{B \to \pm i} \pm ic$ \et             
  &\bt$\xrightarrow{B \to \pm i}\mp i\frac{b}{c}+D\exp(\mp i c(x-A))$\et
    &\bt Type $(B)$ \et                                         \\       
                &               &               &               \\
\hline
\end{tabular}
\caption{General solutions of the equations  
(\ref{eqk1}) and (\ref{eqk0}), and some limiting cases. $A$ and
$B$  are integration constants. The constant
$B$ selects the particular solution of (\ref{eqk1}) in each case.
$D$ is not defined always the same way, but always
represents an arbitrary constant. 
}
\label{res_p_InHu}
\end{table}

Let us now try to further generalize (\ref{mlin}) to higher powers of $m$.
If we try   
\begin{equation}
k(x,m)=k_0(x)+m\,k_1(x)+m^2\,k_2(x)\,,\label{mquad}
\end{equation}
substituting it into (\ref{lhs}) we obtain 
\ba
&&L(m)-L(m+1)                                           
=4 m^3 k_2^2+2 m^2(3 k_1 k_2+3 k_2^2+k'_2)         \nonumber\\
&&\quad\quad+2 m (k_1^2+3 k_1 k_2+2 k_2^2+2 k_0 k_2+k'_1+k'_2)
+\dots\,, \nonumber
\ea
where the dots stand for terms not involving $m$.
Since the coefficients of powers of $m$ must be constant,
from the term in $m^3$ we have $k_2=\hbox{Const}$. From the other terms,
if $k_2\neq 0$ we obtain that both of $k_1$ and $k_0$
have to be constant as well. That is, a case of the
trivial solution $k(x,m)=f(m)$.
The same procedure can be used to show that further generalizations to higher
powers of $m$ give no new solutions \cite{InfHul}.

Let us try now the simplest
generalization of (\ref{mlin}) to inverse powers of $m$. 
Assuming $m\neq 0$, we propose
\begin{equation}
k(x,m)=\frac{k_{-1}(x)}{m}+k_0(x)+m k_1(x)\,.
\label{minv1}
\end{equation}
Substituting into (\ref{lhs}) we obtain 
\ba
L(m)-L(m+1)=\frac{(2m+1)k_{-1}^2}{m^2(m+1)^2}
-2\frac{k_0\,k_{-1}}{m(m+1)}+\frac{(2m+1)k'_{-1}}{m(m+1)}+\dots\,,
\nonumber
\ea
where the dots denote now the right hand side of (\ref{eqLssimp}).
Then, in addition to the equations (\ref{eqk1}) and (\ref{eqk0}) the following
have to be satisfied
\ba
k_{-1}^2=e\,, \quad k_0\,k_{-1}=f\,,\quad  k'_{-1}=g\,,\label{ad_eq_inv1}
\ea
where the right hand side of these equations are constants. 
Is easy to prove that the only non--trivial new solutions 
appear when $k_{-1}(x)=q$, with $q$ non-vanishing constant, 
$k_0(x)=0$ and $k_1(x)$ is not constant. 
We have to consider then
the general solutions of (\ref{eqk1}) for each sign of $a$, shown in 
Table~\ref{res_p_InHu}. The new results are shown
in Table~\ref{sols_m_inv}. In this table, to obtain really 
different new non--trivial solutions, $B$ should be different 
from $\pm 1$ in the case $a>0$, and different from $0$ in the case 
$a=0$, otherwise we would obtain constant particular solutions of
(\ref{eqk1}).

For the case $a<0$, taking $B\to 0$ we recover
the factorization type $(E)$ of Infeld and Hull \cite[eq. $(3.1.7e)$]{InfHul}.
Again, they use this factorization type for particular cases of physical
interest after having made the formal change $\alpha\to i\alpha$
\cite[pp. 46, 47]{InfHul}. The same result is achieved by considering the 
limiting case $B\to 0$ in  $a>0$. For the case $a=0$, taking 
$B\to \infty$ we recover the factorization type $(F)$ (see their equation
$(3.1.7f)$).
For all these solutions of (\ref{lhs}) of type (\ref{minv1})
the  expression for $L(m)$ is  
$L(m)=-a m^2-q^2/m^2+t$, with $t$ an arbitrary real constant, which is
also valid for the case $a=0$.

\begin{table}
\begin{tabular}{|p{.65in}|p{2.1in}|p{.625in}|p{.625in}|p{.5in}|}
\hline
\multicolumn{1}{|c|}{\bt Sign of $a$\et}
        &\multicolumn{1}{|c|}{\bt $k_1(x)$ and limiting cases\et}
             &\multicolumn{1}{|c|}{\bt $k_0(x)$\et}
                  &\multicolumn{1}{|c|}{\bt $k_{-1}(x)$\et}     
                        &\multicolumn{1}{|c|}{\bt Comments\et}  \\
\hline
                &               &               &       &       \\    
\bt$a=c^2>0$\et 
 &\bt\quad$c\,\frac{B\,\sinh(c(x-A))-\cosh(c(x-A))}
   {B\,\cosh(c(x-A))-\sinh(c(x-A))}$\et
        &\bt\quad\quad$0$\et    &\bt\quad$q\in\R$\et    &       \\
                &               &               &       &       \\       
 &\bt\quad$\xrightarrow{B \to 0} c\coth(c(x-A))$\et             
                &\bt\quad\quad$0$\et    &\bt\quad$q\in\R$\et    
        &\bt See text\et                                                 \\
                &               &               &       &       \\
\hline
                &               &               &       &       \\       
\bt$a=0$\et 
    &\bt\quad$\frac{B}{1+B(x-A)}$\et
  &\bt\quad\quad$0$\et  &\bt\quad$q\in\R$\et            &       \\
                &               &               &       &       \\       
        &\bt\quad$\xrightarrow{B \to \infty} \frac{1}{x-A}$\et          
  &\bt\quad\quad$0$\et  &\bt\quad$q\in\R$\et    
        &\bt Type $(F)$\et                                      \\
                &               &               &       &       \\       
\hline
                &               &               &       &       \\        
\bt$a=-c^2<0$\et
  &\bt\quad$-c\,\frac{B\,\sin(c(x-A))+\cos(c(x-A))}
  {B\,\cos(c(x-A))-\sin(c(x-A))}$\et
        &\bt\quad\quad$0$\et    &\bt\quad$q\in\R$\et            &       \\
                &               &               &       &       \\       
 &\bt\quad$\xrightarrow{B \to 0} c\cot(c(x-A))$\et              
        &\bt\quad\quad$0$\et    &\bt\quad$q\in\R$\et    
        &\bt Type $(E)$\et                                      \\
                &               &               &       &       \\
\hline
\end{tabular}
\caption{New solutions of equations (\ref{eqk1}), (\ref{eqk0}) and 
(\ref{ad_eq_inv1}). $A$ is an arbitrary constant. $B$ selects 
the particular solution of (\ref{eqk1}) for each sign of $a$. 
}
\label{sols_m_inv}
\end{table}

It can be checked that further generalizations of (\ref{minv1}) to
higher negative powers of $m$ lead to no new solutions apart from
the trivial one and that of Tables~\ref{res_p_InHu} and~\ref{sols_m_inv}. 

As a consequence, we have obtained \emph{all} possible solutions
of (\ref{lhs}) for $k(x,m)$ if it takes the form of a finite sum of terms 
involving functions of only $x$ times powers of $m$. As a consequence 
of Corollary~\ref{equiv_SI1p_FM} we have found six different, and rather 
general families of Shape--Invariant
potentials in the sense of \cite{Gen} which depend on only one parameter $m$
transformed by translation. These are calculated by means of the formulas
(\ref{ricVSI1p}), (\ref{ricVtilSI1p}), (\ref{idWk}) and (\ref{idRL}).
We show the final results in 
Tables~\ref{sols_k_fin}, \ref{sols_pot_fin_1} and \ref{sols_pot_fin_2}.
We would like to remark here several relations that satisfy the functions 
defined in Table~\ref{sols_k_fin}. In the case $a=c^2$ we have
\ba
f'_{+}=c(1-f_{+}^2)=c(B^2-1)h_{+}^2\,,
\quad\quad h'_{+}=-c f_{+} h_{+}\,,                      \nonumber
\ea
in the case $a=0$, 
\ba
f'_{0}=-B\,f_{0}^2\,,                         
\quad\quad h'_{0}=-B\,f_{0} h_{0}+1\,,                   \nonumber
\ea 
and finally in the case $a=-c^2$,
\ba
f'_{-}=c(1+f_{-}^2)=c(B^2+1)h_{-}^2\,,        
\quad\quad h'_{-}=c f_{-} h_{-}\,,                       \nonumber
\ea
where the prime means derivative respect to $x$
and the arguments are the same as in the 
mentioned table, but have been dropped out for simplicity.

\begin{table}
\begin{tabular}{|p{.8in}|p{2.3in}|p{1.73in}|}
\hline
\multicolumn{1}{|c|}{\bt Sign of $a$\et}
     &\multicolumn{1}{|c|}{\bt$k(x,m)=k_0(x)+m\,k_1(x)$, $L(m)$\et}
      &\multicolumn{1}{|c|}{\bt$k(x,m)=q/m+k_1(x)$, $L(m)$\et}\\
\hline
                &               &                               \\    
\bt\quad$a=c^2>0$\et    
&\bt\quad$\frac{b+ma}{c} f_{+}(x,A,B,c)+D h_{+}(x,A,B,c)$\et
                    &\bt\quad$\frac{q}{m}+m c f_{+}(x,A,B,c)$\et        \\
                &               &                               \\
                &\bt\quad$-c^2 m^2-2 b m +t$\et
                        &\bt\quad$-c^2 m^2-\frac{q^2}{m^2}+t$\et        \\
                &               &                               \\
\hline
                &               &                               \\       
\bt\quad$a=0$\et 
                &\bt\quad$b\,h_0(x,A,B)+(m B+D) f_0(x,A,B)$\et
                   &\bt\quad$\frac q m +m B f_0(x,A,B)$\et      \\
                &               &                               \\
                &\bt\quad$-2 b m +t$\et
                        &\bt$\quad-\frac{q^2}{m^2}+t$\et                \\
                &               &                               \\         
\hline
                &               &                               \\        
\bt\quad$a=-c^2<0$\et
                &\bt\quad$\frac{b+ma}{c}f_{-}(x,A,B,c)+D h_{-}(x,A,B,c)$\et
                 &\bt\quad$\frac{q}{m}-m c f_{-}(x,A,B,c)$\et   \\
                &               &                               \\
                &\bt\quad$c^2 m^2-2 b m +t$\et
                        &\bt\quad$c^2 m^2-\frac{q^2}{m^2}+t$\et \\
                &               &                               \\      
\hline
\end{tabular}
\begin{tabular}{|p{2.4in}p{2.6in}|}
\hline
\multicolumn{1}{|c}{}
                        &\multicolumn{1}{c|}{}                  \\
\bt where \et   &                                               \\
\bt\ \ \ $f_{+}(x,A,B,c)=\frac{B\,\sinh(c(x-A))-\cosh(c(x-A))}
{B\,\cosh(c(x-A))-\sinh(c(x-A))}$\et    
 &\bt\quad$h_{+}(x,A,B,c)
   =\frac 1 {B\,\cosh(c(x-A))-\sinh(c(x-A))}$\et                \\
                &                                               \\
\bt\ \ \ $f_{0}(x,A,B)=\frac{1}{1+B(x-A)}$\et   
 &\bt\quad$h_{0}(x,A,B)=\frac{\frac B 2 (x-A)^2+x-A}{1+B(x-A)}$\et      \\
                &                                               \\
\bt\ \ \ $f_{-}(x,A,B,c)=\frac{B\,\sin(c(x-A))+\cos(c(x-A))}
        {B\,\cos(c(x-A))-\sin(c(x-A))}$\et      
 &\bt\quad$h_{-}(x,A,B,c)=\frac{1}
        {B\,\cos(c(x-A))-\sin(c(x-A))}$\et                      \\      
                &                                               \\
\hline
\end{tabular}
\caption{General solutions for the two forms of $k(x,m)$  
(\ref{mlin}) and (\ref{minv1}). 
$A$, $B$, $D$, $q$ and $t$ are arbitrary constants. The constant
$B$ selects the particular solution of (\ref{eqk1}) for each sign of $a$. 
The constant $b$ is that of (\ref{eqk0}).
}
\label{sols_k_fin}
\end{table}

\begin{table}
\begin{tabular}{|p{.7in}|p{4.33in}|}
\hline
\multicolumn{1}{|c|}{\bt Sign of $a$\et}
        &\multicolumn{1}{|c|}{\bt$V(x,m)-d$, $\til V(x,m)-d$, $R(m)$ 
when $k(x,m)=k_0(x)+m k_1(x)$\et}\\
\hline
                &                                               \\    
\bt$a=c^2>0$\et 
&\bt\quad$\frac{(b+ma)^2}{a} f_{+}^2
        +\frac{D}{c}(2(b+ma)+a) f_{+} h_{+}
        +(D^2-(B^2-1)(b+ma))h_{+}^2$\et                         \\
                &                                               \\
 &\bt\quad$\frac{(b+ma)^2}{a} f_{+}^2
        +\frac{D}{c}(2(b+ma)-a) f_{+} h_{+}
        +(D^2+(B^2-1)(b+ma))h_{+}^2$\et                         \\
                &                                               \\
 &\bt\quad$R(m)=L(m)-L(m+1)=2(b+ma)+a$\et               \\
                &                                               \\
\hline
                &                                               \\
\bt$a=0$\et     
  &\bt\quad$b^2 h_0^2+(D+mB)(D+(m+1)B)f_0^2+2 b 
        (D+(m+\frac 1 2)B)f_0 h_0-b$\et                         \\
                &                                               \\
 &\bt\quad$b^2 h_0^2+(D+mB)(D+(m-1)B)f_0^2+2 b 
        (D+(m-\frac 1 2)B)f_0 h_0+b$\et                         \\
                &                                               \\
 &\bt\quad$R(m)=L(m)-L(m+1)=2 b$\et                     \\
                &                                               \\
\hline
                &                                               \\
\bt$a=-c^2<0$\et        
&\bt\quad$-\frac{(b+ma)^2}{a} f_{-}^2
        +\frac{D}{c}(2(b+ma)+a) f_{-}h_{-}
        +(D^2-(B^2+1)(b+ma))h_{-}^2$\et                         \\
                &                                               \\
 &\bt\quad$-\frac{(b+ma)^2}{a} f_{-}^2
        +\frac{D}{c}(2(b+ma)-a) f_{-}h_{-}
        +(D^2+(B^2+1)(b+ma))h_{-}^2$\et                         \\
                &                                               \\
 &\bt\quad$R(m)=L(m)-L(m+1)=2(b+ma)+a$\et               \\
                &                                               \\
\hline
\end{tabular}
\begin{tabular}{|p{5.2in}|}
\hline
\multicolumn{1}{|c|}{}
                                                                \\
\bt where \qquad$f_{+}=f_{+}(x,A,B,c)$,\quad $f_{0}=f_{0}(x,A,B)$,
\quad $f_{-}=f_{-}(x,A,B,c)$\et                                 \\      
\bt\qquad\qquad\quad$h_{+}=h_{+}(x,A,B,c)$,\quad $h_{0}=h_{0}(x,A,B)$,
\quad $h_{-}=h_{-}(x,A,B,c)$ 
        \quad are defined as in Table~\ref{sols_k_fin}\et       \\
                                                                \\
\hline
\end{tabular}
\caption{ Shape--Invariant potentials which depend on one
parameter $m$ transformed by traslation, when $k(x,m)$ is of the form
(\ref{mlin}). 
$A$, $B$, and $D$  are arbitrary constants. 
The constant
$B$ selects the particular solution of (\ref{eqk1}) for each
sign of $a$. The constant $b$ is that of (\ref{eqk0}).
The Shape Invariance condition $\til V(x,m)=V(x,m-1)+R(m-1)$
is satisfied in all cases.}
\label{sols_pot_fin_1}
\end{table}

\begin{table}
\begin{tabular}{|p{.7in}|p{4.33in}|}
\hline
\multicolumn{1}{|c|}{\bt Sign of $a$\et}
        &\multicolumn{1}{|c|}{\bt$V(x,m)-d$, $\til V(x,m)-d$, $R(m)$ 
when $k(x,m)=q/m+m k_1(x)$\et}\\
\hline
                &                                               \\    
\bt$a=c^2>0$\et 
&\bt\quad$\frac{q^2}{m^2}+m^2 c^2+2 q c f_{+}
        -m(m+1)c^2 (B^2-1)h_{+}^2$\et                           \\
                &                                               \\
 &\bt\quad$\frac{q^2}{m^2}+m^2 c^2+2 q c f_{+}
        -m(m-1)c^2 (B^2-1)h_{+}^2$\et                           \\
                &                                               \\
 &\bt\quad$R(m)=L(m)-L(m+1)
        =\frac{q^2}{(m+1)^2}-\frac{q^2}{m^2}+(2 m+1)c^2$\et     \\
                &                                               \\
\hline
                &                                               \\
\bt$a=0$\et     
  &\bt\quad$\frac{q^2}{m^2}+2 q B f_0+m(m+1) B^2 f_0^2$\et      \\
                &                                               \\
  &\bt\quad$\frac{q^2}{m^2}+2 q B f_0+m(m-1) B^2 f_0^2$\et      \\
                &                                               \\
 &\bt\quad$R(m)=L(m)-L(m+1)
                =\frac{q^2}{(m+1)^2}-\frac{q^2}{m^2}$\et        \\
                &                                               \\
\hline
                &                                               \\
\bt$a=-c^2<0$\et        
&\bt\quad$\frac{q^2}{m^2}-m^2 c^2-2 q c f_{-}
        + m(m+1) c^2 (B^2+1)h_{-}^2$\et                         \\
                &                                               \\
 &\bt\quad$\frac{q^2}{m^2}-m^2 c^2-2 q c f_{-}
        + m(m-1) c^2 (B^2+1)h_{-}^2$\et                         \\
                &                                               \\
 &\bt\quad$R(m)=L(m)-L(m+1)
        =\frac{q^2}{(m+1)^2}-\frac{q^2}{m^2}-(2 m+1)c^2$\et     \\
                &                                               \\
\hline
\end{tabular}
\begin{tabular}{|p{5.2in}|}
\hline
\multicolumn{1}{|c|}{}
                                                                \\
\bt where \qquad$f_{+}=f_{+}(x,A,B,c)$,\quad $f_{0}=f_{0}(x,A,B)$,
\quad $f_{-}=f_{-}(x,A,B,c)$\et                                 \\      
\bt\qquad\qquad\quad$h_{+}=h_{+}(x,A,B,c)$,\quad $h_{0}=h_{0}(x,A,B)$,
\quad $h_{-}=h_{-}(x,A,B,c)$ 
        \quad are defined as in Table~\ref{sols_k_fin}\et       \\
                                                                \\
\hline
\end{tabular}
\caption{Shape--Invariant potentials which depend on one
parameter $m$ transformed by traslation, when $k(x,m)$ is of the form
(\ref{minv1}). 
$A$, $B$, $D$ and $q$ are arbitrary constants. 
The constant
$B$ selects the particular solution of (\ref{eqk1}) for each
sign of $a$. The constant $b$ is that of (\ref{eqk0}).
The Shape Invariance condition $\til V(x,m)=V(x,m-1)+R(m-1)$
is satisfied in all cases.
}
\label{sols_pot_fin_2}
\end{table}

\section{Conclusions and outlook}

After a quick review of basic concepts in the theory of 
factorizable Hamiltonians
and Supersymmetric Quantum Mechanics, 
we have carefully analyzed the equivalence between 
a generalization of the Factorization Method given in \cite{InfHul}
as to allow the relevant parameters to change in an arbitrary way, 
and the Shape Invariant potentials theory.

We have treated the particularly simple but important case of only
one parameter subject to translations, that is, the kind of problems
treated by Infeld and Hull in their classic paper. To do that, we 
have considered the general solutions of certain Riccati equation with
constant coefficients rather than particular ones. 
As a result, we have obtained more general classes of factorizable problems
(resp. Shape Invariant partner potentials) than the ones appearing in 
\cite{InfHul}.

On the other hand, the bridge beetween Shape Invariance and factorizable
problems has been established more clearly. To this respect, we would like
to remark that in the interesting paper \cite[Sec. \bf{VI}]{CoopGinKha}
a classification of several solutions to the Shape Invariance condition 
(\ref{SIGed1p}) is given. Comparing their \emph{ans\"astze} for the superpotential
(6.8) with the one proposed by Infeld and Hull, reproduced here as (\ref{minv1}),
is even more clear the relation between both approaches. In both of them, the
solutions can be generalized simply considering the general solutions of a
Riccati equation, as we have shown in this article.

But what is even more important is that the use of the properties of the 
Riccati equation provides a great insight in order to study still unsolved
problems as the one suggested in the end of \cite[Sec. \bf{VI}]{CoopGinKha}.
That is the subject of another article \cite{CarRamtres}.

Finally, we would like to note, 
since \cite{InfHul} is a very referenced and used paper, 
that we have detected one missprint there which may 
produce later unaccurate results. 
In the expression of the factorization of general Type B of 
\cite[page 36]{InfHul}, $k(x,m)$ should be $d\exp(a x)-a(m+c)$
instead of $d\exp(a x)-m-c$, according to
their notation. This missprint is reproduced in their final 
table of factorizations, page 67. However, the function $r(x,m)$ 
they give for that $k(x,m)$ is correct.

\bigskip

{\parindent 0cm{\Large \bf Acknowledgements.}} 

\medskip
One of the authors (A.R.) thanks the Spanish Ministerio de 
Educaci\'on y Cultura for a FPI grant, research project 
PB96--0717. Support of Spanish
DGES (PB96--0717) is also acknowledged.


\begin{thebibliography}{xx}

\bibitem{CarMarNas}  
Cari\~nena J.F., Marmo G., and Nasarre J.,
{\it  The nonlinear superposition principle 
and the Wei--Norman method\/},
Int. J. Mod. Phys. {\bf 13}, 3601--27 (1998). 

\bibitem{CarMarPerRan}
Cari\~nena J.F.,  Marmo G.,  Perelomov A.M. and  Ra\~nada M.F.,
{\em Related operators  and exact solutions  of Schr\"odinger equations\/},
Int. J. Mod. Phys. {\bf A 13}, 4913--29 (1998).

\bibitem{CarRam}  
Cari\~nena J.F. and Ramos A.,
{\it Integrability of the Riccati equation 
from a group theoretical viewpoint\/},
Int. J. Mod. Phys. {\bf A 14}, 1935--51 (1999).

\bibitem{CarRamtres}  
Cari\~nena J.F. and Ramos A.,
{\it Shape invariant potential depending on $n$ parameters
transformed by translation\/},
DFTUZ Preprint 99/09, (1999). Submitted.

\bibitem{CarRamGra}  
Cari\~nena J.F., Grabowski J. and Ramos A.,
{\it Reduction of time--dependent systems admitting 
a superposition principle\/},
DFTUZ Preprint 99/07, (1999). Submitted.

\bibitem{CoopGinKha}
Cooper F., Ginocchio J.N. and Khare A.,
{\it Relationship between supersymmetry and 
solvable potentials},
Phys. Rev. {\bf 36 D}, 2458--73 (1987). 

\bibitem{Dav} 
Davis H.T.,
{\it Introduction to Nonlinear Differential and 
Integral Equations\/},
(Dover, New York, 1962).

\bibitem{Gen}
Gendenshte\"{\i}n L.\'E., 
{\em Derivation of exact spectra of
the Schr\"odinger equation by means of supersymmetry},
JETP Lett. {\bf 38}, 356--9 (1983). 

\bibitem{GenKriv}
Gendenshte\"{\i}n L.\'E. and Krive I.V., 
{\em  Supersymmetry in quantum mechanics},
Soviet Phys. Usp. {\bf 28}, 645--66 (1985). 

\bibitem{Hum68}
Humi M.,
{\em Extension of the Factorization Method\/}, 
J. Math. Phys. {\bf 9}, 1258--65 (1968). 

\bibitem{Hum70}
Humi M.,
{\em New types of factorizable equations\/}, 
Proc. Camb. Phil. Soc. {\bf 68}, 439--46 (1970). 

\bibitem{Hum86}
Humi M.,
{\em Factorization of systems of differential equations\/}, 
J. Math. Phys. {\bf 27}, 76--81 (1986). 

\bibitem{Hum87}
Humi M.,
{\em Novel types of factorisable systems of differential equations\/}, 
J. Phys. A: Math. Gen. {\bf 20}, 1323--31 (1987). 

\bibitem{InfHul}
Infeld L. and Hull T.E., 
{\em The Factorization Method\/},
Rev. Mod. Phys.  {\bf 23}, 21--68 (1951). 

\bibitem{LS}   
Lie S. and Scheffers  G.,
{\it Vorlesungen \"uber continuierlichen Gruppen mit 
geometrischen und anderen Anwendungen\/},
(Teubner, Leipzig, 1893).

\bibitem{MonSal} 
Montemayor R. and Salem L.D., 
{\it Supersymmetry shape invariance and solubility in quantum mechanics\/},
Phys. Rev. {\bf A 40}, 2170--2173 (1987). 
 
\bibitem{Mur}   
Murphy G.M.,
{\it Ordinary Differential equations and their solutions\/}, 
(Van Nostrand, New York, 1960).

\bibitem{Sch1}
Schr\"odinger E.,
{\em A method of determining quantum--mechanical 
eigenvalues and eigenfunctions\/},
Proc. Roy. Irish Acad. A {\bf XLVI}, 9--16 (1940). 

\bibitem{Sch2}
Schr\"odinger E.,
{\em Further studies on solving eigenvalue problems by factorization\/},
Proc. Roy. Irish Acad. A {\bf XLVI}, 183--206 (1941). 

\bibitem{Sch3}
Schr\"odinger E.,
{\em The factorization of the hypergeometric equation\/},
Proc. Roy. Irish Acad. A {\bf XLVII}, 53--54 (1941). 

\bibitem{ShnWin}
Shnider S. and Winternitz P.,
{\em Classification of Systems of ordinary differential equations with 
superposition principles\/},
J. Math. Phys.  {\bf 25}, 3155--65 (1984).

\bibitem{Sta}
Stahlhofen A.,
{\it The Riccati equation as a 
common basis for Su\-per\-sym\-me\-tric
Quantum Mechanics and the Factorization Method \/}, 
Preprint Duke University (1988). 

\bibitem{Sta2}
Stahlhofen A.,
{\it Remarks on the equiva\-len\-ce bet\-ween the Sha\-pe--In\-va\-rian\-ce
condition and the factorisation condition\/},
J. Phys. A: Math. Gen. {\bf 22}, 1053--8 (1989). 

\bibitem{PW1}  
Winternitz P.,
{\it Lie groups and solutions of nonlinear differential equations\/},
in Nonlinear Phenomena, K.B. Wolf Ed., Lecture Notes in Physics {\bf 189},
(Springer-Verlag, N.Y., 1983).

\bibitem{PW2}  
Winternitz P.,
{\it Comments on  superposition rules for nonlinear coupled first order
differential  equations\/},
J. Math. Phys. {\bf 25}, 2149--50 (1984).

\bibitem{Witten81}
Witten E.,
{\it Dynamical breaking of Supersymmetry\/},
Nucl. Phys. {\bf B 188}, 513--54 (1981). 

\end{thebibliography}
\end{document}